\documentclass[draft,a4paper]{article}

\usepackage[figuresright]{rotating}
\usepackage{graphicx}

\newcommand{\nucl}[3]{Nucl. Phys. \textbf{#1} ({#2}) {#3}.}
\newcommand{\rmp}[3]{Rev. Mod. Phys. \textbf{#1} ({#2}) {#3}.}
\newcommand{\prd}[3]{Phys. Rev. \textbf{D{#1}} ({#2}) {#3}.}
\newcommand{\plb}[3]{Phys. Lett. \textbf{B{#1}} ({#2}) {#3}.}
\newcommand{\prep}[3]{Phys. Rep. \textbf{{#1}} ({#2}) {#3}.}
\newcommand{\ppnp}[3]{Prog. Part. Nucl. Phys. \textbf{{#1}} ({#2}) {#3}.}
\renewcommand{\Box}{\partial^2}

\begin{document}

\title{Classical Solutions of SU(2) and SU(3) Pure Yang-Mills Theories and
       Heavy Quark Spectrum}
\author{O. Oliveira\thanks{\textit{email}: orlando@teor.fis.uc.pt},
        R. A. Coimbra\thanks{\textit{email}: rita@teor.fis.uc.pt} \\
        Centro de {F\'\i}sica Computacional \\
        Departamento de F{\'\i}sica \\
        Universidade de Coimbra \\
        3004-516 Coimbra \\
        Portugal }

\maketitle

\begin{abstract}
In this paper we compute classical Minkowsky spacetime solutions of pure SU(2)
and SU(3) gauge theories, in Landau gauge. The solutions are regular 
everywhere except at the origin and/or infinity, are characterized by a four
momentum $k$ such that $k^2 \, = \, 0$ and resemble QED configurations.
The classical solutions suggest a particle-independent description of
hadrons, similarly to the Atomic and Nuclear energy levels, which is able to 
reproduce the heavy quarkonium spectrum with a precision below $10 \%$.
Typical errors in the theoretical mass prediction relative
to the measured mass being of the order of 2-4\%.
\end{abstract}

\section{Introduction and Motivation}

For nonabelian gauge theories the classical field equations are nonlinear 
partial differential equations. Finding solutions of the classical field
equations has proved to be quite a challenge and we still do not know its
general solution. The usual approach to classical gauge configurations is to
looks for ansatz that simplify the mathematical problem at hand. The ansatz
should allow the computation of solutions of the partial differential 
equations and, simultaneously, should provide a proper description of the
field's dynamics.

For pure gauge SU(2) theory several classical configurations are known.
They have been computed using different ansatz, see \cite{Ac79} for a review.
For SU(3), classical configurations can be built from SU(2) solutions. 
In principle, the SU(3) gauge dynamics goes beyond SU(2) classical 
configurations and, in order to understand its dynamics, one should try to
enlarge the set of known solutions as much as possible.

The interest on the classical gauge configurations goes beyond classical
field theory. Indeed, the classical solutions of a field theory are a first
step towards the understanding of the associated quantum theory. 
For QCD, classical configurations built from SU(2) configurations, in 
particular the instanton \cite{Ho79,Ho81,Po77,ScSh98,Di02}, are used
to address different aspects of hadronic phenomenology. The large number of 
studies involving instantons shows that classical gauge configurations have
a role to play in the associated quantum gauge theory. Hopefully, it would be
interesting if these configurations could provide hints about two of the main
open problems of strong interactions, namely quark confinement in hadrons
and/or chiral symmetry breaking mechanism. 

In this paper we report on new Minkowsky spacetime solutions for pure gauge 
classical SU(2) and SU(3) theories. 
The Landau gauge solutions were obtained after 
using a technique inspired on Cho's work 
\cite{Ch991,Ch992,ChLe99,FaNi99} to write the Yang-Mills fields.
The generalized Cho-Faddeev-Niemi ansatz, combined with a spherical-like basis
in color space, reduces the nonlinear field equations to linear abelian-like
equations and replaces the Landau gauge condition, a linear gauge condition,
by a set of coupled equations.

For SU(2), the gauge fields are functions of a vector field $\hat{A}_\mu$ and
a scalar field $\theta_2$ and the ansatz reproduces the original 
Cho-Faddeev-Niemi ansatz. For SU(3), besides $\hat{A}_\mu$ and $\theta_2$
an extra vector field $C_\mu$ is required. The relation between 
$\hat{A}_\mu$, $\theta_2$, $C_\mu$ and the gluon fields is nonlinear. 
For the ansatz discussed in the paper, the classical field equations become
Maxwell equations for $\hat{A}_\mu$ and $C_\mu$ but the Euler-Lagrange
equations don't provide information on $\theta_2$. 
For pure gauge theories, the classical action is independent of $\theta_2$.
It is the gauge fixing condition which restricts $\theta_2$.
This seems to suggest that $\theta_2$ is not a dynamical field and, 
therefore, that it can be chosen arbitrarily. However, if one considers matter
fields the action is a functional of $\partial_\mu \theta_2$. Instead of
fixing $\theta_2$ \textit{a priori}, we choose to compute this scalar function
solving simultaneously the classical equations of motion and the gauge fixing
condition. 

The classical field configurations considered in this paper have finite
action and energy. They are regular everywhere, except at the origin 
and/or at infinity, and are characterized by a four-vector $k$ such that 
$k^2 \, = \, 0$, like in the solutions of the linearized field equations. 
Each four-vector $k$ identifies linear combinations of solutions of the type
\begin{displaymath}
  a  \, e^{\pm i kx} ~  + ~  b \, e^{\pm  kx} \, .
\end{displaymath}
The first components are the usual free field solutions and are eigenfunctions
of $i \partial_\mu$ with real eigenvalues. The second components are also
eigenfunctions of $i \partial_\mu$ but with pure imaginary eigenvalues.
In the general solution of the Euler-Lagrange equations, the vector fields 
$\hat{A}_\mu$ and $C_\mu$ are linear combinations of the pure complex 
exponential functions. The scalar $\theta_2$ is given by a combination of both
type of exponential functions.
The ansatz allows the computation of classical vacuum configurations, fields 
that share the same general properties as those just described.

After discussing classical solutions of the field equations, we investigate 
the properties of quarks in the background of the classical gauge 
configurations. We show that the quark problem can be reduced to three 
electromagnetic-type problems and, that from the point of view of these 
electromagnetic problems, not all the couplings are atractive couplings - the 
maximum number of atractive couplings being two. 
Moreover, if one look for eigenfunctions of $i \partial_0$, it turns out that
they are the gauge invariant combinations which represent mesons
${\overline \psi}^a \psi^a$ and baryons $ \epsilon_{abc} \psi^a \psi^b \psi^c$
in the quark model. The third color component of $\psi$ and the product 
$\psi^1 \, \psi^2$ of the color components are also 
eigenfunctions of $i \partial_0$, but are  not gauge invariant wave 
functions. Motivated by this observation, we try to understand the heavy 
quarkonium meson spectrum, i.e. charmonium and bottomonium, from the point of 
view of quarks propagating in the background of classical configurations. 
Assuming that the spectrum is a combination of two hidrogenium like spectrum
and identifying physical states as eigenstates of $i \partial_0$, we are able
to reproduce the quarkonium meson spectrum below 10$\%$ error for all mesons,
with typical errors below 4$\%$ for the meson masses. Note that this study
of the particles spectrum is a first order approximation to the 
complete problem. The level of accuracy achieve with such a simple picture
of an hadron suggests that it is a good starting point for a more detailed
calculation.

The paper is organized has follows. In section 2 we discuss the construction
of the ansatz for SU(2) and SU(3) pure gauge fields. In section 3 we compute
Minkowsky space-time SU(2) solutions in Landau gauge. Section 4 repeats the 
previous section but for SU(3) gauge theory.
In section 5 we discuss the coupling of quarks to the classical
SU(3) configurations and in section 6 discuss the relevance of this 
configurations for the heavy quarkonium spectrum. In section 7 we give the
conclusions. The appendix contains material used along the paper.

\section{Gauge Fields for SU(2) and SU(3)}

For $SU(N)$ gauge theories the lagrangian reads
\begin{equation}
 \mathcal{L} \, = \, - \frac{1}{4} \, F^a_{\mu\nu} \, F^{a \, \mu\nu}
\end{equation}
where
\begin{equation}
 F^a_{\mu\nu} \, = \, \partial_\mu A^a_\nu \, - \,
                      \partial_\nu A^a_\mu  \, - \,
                      g f_{abc} A^b_\mu A^c_\nu
 \mbox{ } ,
\end{equation}
and $A^a_\mu$ are the gluon fields. 

Let $n^a$ be a covariant constant real scalar field in the adjoint 
representation. From the definition it follows that
\begin{equation}
 D_\mu \, n ^a \, = \, 
    \partial_\mu n^a \, + \, i g \left( F^b \right)_{ac} A^b_\mu n^c
  \, = \, 0 \, ; \label{Dn1}
\end{equation}
the generators of the adjoint representation are 
$\left( F^b \right)_{ac} \, = \, -i \, f_{bac} \,$. 
Given a gluon field it is always possible to solve the above equations for 
the scalar field $n$. The set of equations are linear partial differential
equations for $n$. In general, the solution of (\ref{Dn1}) is not 
uniquely defined unless boundary conditions are provided. Solving (\ref{Dn1})
is similar to solve the Laplace equation, where the uniqueness theorem 
requires boundary conditions. From a formal point of view, the set of
equations (\ref{Dn1}), together with appropriate boundary conditions, defines
a mapping from the gluon field $A^a_\mu$ to $n^a$.

Can we say something about the inverse mapping? From the point of view
of $A^a_\mu$, equation (\ref{Dn1}) is a linear equation. This
observation may suggest that given a scalar field $n$, the solution of 
equations  (\ref{Dn1}) is a unique gluon field $A^a_\mu$. 
However, we will see that, in general, $n$ does not uniquely determines 
the gluon field, although it helps in reducing the number of independent 
fields that should be considered when writing $A^a_\mu$. 

Let us discuss the mapping $n \, \longrightarrow \, A^a_\mu$.
From (\ref{Dn1}) it follows
\begin{equation}
    n \partial_\mu n \, = \, \frac{1}{2} \, \partial_\mu n^2 \, = \, 0,
   \label{t1}
\end{equation}
and one can always choose $n^2 \, = \, 1$. Let us write the gluon field as
\begin{equation}
 A^a_\mu \, = \, n^a \hat{A}_\mu \, + \, X^a_\mu \, ,
\end{equation}
where the field $X$ is orthogonal to $n$ in the sense
\begin{equation}
 n \cdot X_\mu \, = \, \sum\limits^{}_{a} \, n^a \, X^a_\mu \, = \, 0 \, .
\end{equation}
To proceed we multiply (\ref{Dn1}) by  $\left( F^d \right)_{ea} n^d$. For
SU(2) theory one has the following relation
\begin{equation}
       f_{abc} \, f_{dec} \, = \, 
             \delta_{ad} \delta_{be} \, - \,
                                              \delta_{ae} \delta_{bd} \, .
\end{equation}
For SU(3) we use
\begin{equation}
       f_{abc} \, f_{dec} \, = \, 
            \frac{2}{3} \, \left( \delta_{ad} \delta_{be} \, - \,
                                              \delta_{ae} \delta_{bd} \right)
          \, + \,
        \left( d_{adc} d_{bec} \, - \, d_{bdc} d_{aec} \right)
\end{equation}
where
\begin{equation}
 d_{abc} \, = \,
   \frac{1}{4}
   \mbox{Tr} \left( \lambda^a \, \left\{ \,\lambda^b \, , \,
                                           \lambda^c \, \right\}
             \right) \, ,
\end{equation}
$\lambda^a$ being the Gell-Mann matrices for SU(3).
Solving for the gauge fields we get, after some algebra,
\begin{eqnarray}
 f_{eda} \, n^d \, \partial_\mu n^a \, - \, 
         \Lambda \,
         g \, X^e_\mu \, - \,
         g \left( d_{ebh} \, d_{dch} \, - \, d_{dbh} \, d_{ech} \right) \,
                      X^b_\mu \, n^c \, n^d \, = \, 0 \, .
 \label{eqn1}
\end{eqnarray}
In (\ref{eqn1}) $\Lambda$ is defined as follows
\begin{equation}
   \Lambda \, = \, \left\{ \begin{array}{lcl}
                  \frac{2}{3} \, , & & \mbox{SU(3)} \, , \\
                  1 \, ,           & & \mbox{SU(2)} \, .
              \end{array} \right.
\end{equation}
Equation (\ref{eqn1}) suggest the following form for $X^a_\mu$,
\begin{equation}
X^a_\mu \, = \, \frac{1}{\Lambda \,g} \, 
          f_{abc} \, n^b \, \partial_\mu n^c
 \, + \, Y^a_\mu
\end{equation}
with $Y^a_\mu$ verifying the constraint
\begin{equation}
   n \cdot Y_\mu \, = \, 0 \, 
\end{equation}
for SU(3) Yang-Mills theory and
\begin{equation}
   Y_\mu \, = \, 0
\end{equation}
for SU(2) gauge group.

In terms of $\hat{A}_\mu$, $n^a$ and $Y^a_\mu$ the gauge fields are given by
\begin{equation}
 A^a_\mu \, = \, \hat{A}_\mu n^a \, + \,
                 \frac{1}{\Lambda \, g} \, f_{abc} \,
                              n^b \, \partial_\mu n^c \, + \,
                 Y^a_\mu
 \, ; \label{A}
\end{equation}
with $n$ and $Y$ verifying the constraints
\begin{eqnarray}
 & & n \cdot Y_\mu \, = \, 0 \, ,  \label{nY} \\
 & & D_\mu n^a \, = \, 0  . \label{Dn}
\end{eqnarray}

Let us consider the gauge transformation properties of $n$, $Y$ and $\hat{A}$. 
The field $n$ is, by definition, covariant constant. It follows that
$ -i \, \left( F^c \right)_{ab} \, n^c \, \partial_\mu n^b $ 
belongs to the adjoint representation of the gauge group. 
Demanding that $Y$ is in the adjoint representation of the group, 
$\hat{A}_\mu$ transforms under the gauge group as follows
\begin{equation}
   \hat{A}_\mu \, \longrightarrow \, \hat{A}_\mu \, + \,
                                     \frac{1}{g} \, 
                                         n \cdot \partial_\mu \omega
                                        \, . 
  \label{Atrans}
\end{equation}
Note that for constant $n$, (\ref{Atrans})  mimics the transformation of an 
abelian field. Constraints (\ref{nY}) and (\ref{Dn}) are scalars under gauge 
transformations and parameterization (\ref{A}) provides a gauge 
invariant decomposition of the gluon field $A^a_\mu$.

If (\ref{A}), (\ref{nY}) and (\ref{Dn}) define a complete parameterization
of the gluon fields, the total number of independent fields on both sides of
(\ref{A}) should be the same. Certainly, the counting of field components
is larger on the r.h.s of (\ref{A}) compared to the l.h.s. 
However, the counting of the number of independent fields is not obvious, 
specially in what concerns $n$, $\hat{A}$ and $Y$, and will not be discussed
in this paper. Instead, we will proceed looking at solutions of the
Euler-Lagrange equations by exploring the ansatz defined in this section.

\section{Classical SU(2) Gauge Theory}

In order to solve classical pure SU(2) gauge theory we consider the
spherical basis in color space defined as
\begin{equation}
e_1 \, = \, \left( \begin{array}{c}
                 s_1 \, c_2 \\
                 s_1 \, s_2 \\
                 c_1
                 \end{array}
          \right) \, ,
 \hspace{1cm}
e_2 \, = \, \left( \begin{array}{c}
                 c_1 \, c_2 \\
                 c_1 \, s_2 \\
                 - s_1
                 \end{array}
          \right) \, ,
 \hspace{1cm}
e_3 \, = \, \left( \begin{array}{c}
                 s_2 \\
                 - c_2 \\
                 0
                 \end{array}
          \right) \, ,
\end{equation}
where $s_i \, = \, \sin \theta_i$, $c_i \, = \, \cos \theta_i$ and
$\theta_1$ and $\theta_2$ are functions of spacetime.
For SU(2), $Y^a_\mu \, = \, 0$ and after identifying $n$ with $e_1$
the gauge fields become
\begin{eqnarray}
  &  A^1_\mu \, = & s_1 \, c_2 \, \left( \hat{A}_\mu \, - \,
                                         \frac{1}{g} \, c_1 \, 
                                               \partial_\mu \theta_2 
                                  \right) \, - \,
             \frac{1}{g} \, s_2 \, \partial_\mu \theta_1  \, , \\
  &  A^2_\mu \, = & s_1 \, s_2 \, \left( \hat{A}_\mu \, - \,
                                          \frac{1}{g} \, c_1 \, 
                                              \partial_\mu \theta_2 
                                  \right) \, + \,
             \frac{1}{g} \, c_2 \, \partial_\mu \theta_1  \, , \\
  &  A^3_\mu \, = & \hat{A}_\mu \, c_1 \, + \,
                        \frac{1}{g} \, s^2_1 \, \partial_\mu \theta_2 \, = \,
                  c_1 \, \left( \hat{A}_\mu \, - \,
                          \frac{1}{g} \, c_1 \, \partial_\mu \theta_2 
                  \right) \, + \,  \frac{1}{g} \, \partial_\mu \theta_2 \, ,
\end{eqnarray}
the gluon field tensor is given by
\begin{equation}
 F^a_{\mu \nu} \, = \, n^a \, \mathcal{F}_{\mu \nu} \, ,
\end{equation}
where
\begin{equation}
 \mathcal{F}_{\mu \nu} \, = \, \partial_\mu \mathcal{A}_\nu \, - \,
                               \partial_\nu \mathcal{A}_\mu \, ,
 \hspace{1cm}
 \mathcal{A}_\mu \, = \, \hat{A}_\mu \, - \,
                          \frac{1}{g} \, c_1 \, \partial_\mu \theta_2 
\end{equation}
and the classical equations of motion
\begin{equation}
   n^a \, \partial_\nu \mathcal{F}^{\mu \nu} \, = \, 0.
\end{equation}
The structure of $n$ reduces the last set of equations to
\begin{equation}
   \partial_\nu \, \mathcal{F}^{\mu \nu} \, = \, 0 \, ,
   \label{su2fieldeq} 
\end{equation}
i.e. the ansatz makes classical pure SU(2) gauge theory
equivalent to Maxwell theory. That the equivalence between classical
SU(2) gauge theory and an abelian theory is not perfect can be seen
by looking at the classical hamiltonian density
\begin{eqnarray}
 & \mathcal{H} & = \, F^{a, \beta 0} \, \partial^0 A^a_\beta  \, - \,
                      \mathcal{L} \nonumber \\
 &             & = \, \mathcal{F}^{\beta 0} \, \partial^0 \mathcal{A}_\beta  
                       \, -  \, \left( - \frac{1}{4} \mathcal{F}^{\mu \nu}
                                                     \mathcal{F}_{\mu \nu} 
                                \right)
                       \, + \, \mbox{additional terms}
\end{eqnarray}
and at the classical spin tensor
\begin{eqnarray}
 & \mathcal{S}^{\alpha \beta} & = \, F^{a, \beta 0} \, A^{a, \alpha} \, - \,
                           F^{a, \alpha 0} \, A^{a, \beta} \nonumber \\
 &  & = \, \mathcal{F}^{\beta 0} \, \mathcal{A}^{\alpha} \, - \,
           \mathcal{F}^{\alpha 0} \, \mathcal{A}^{ \beta} \, + \,
           \frac{\cos \theta_1}{g} \left(
                   \mathcal{F}^{\beta 0} \, \partial^{\alpha} \theta_2 \, - \,
                   \mathcal{F}^{\alpha 0} \, \partial^{ \beta} \theta_2
                                    \right)   \nonumber \\
 &  & = \, \mathcal{F}^{\beta 0} \, \hat{A}^{\alpha} \, - \,
           \mathcal{F}^{\alpha 0} \, \hat{A}^{ \beta} \, .
\end{eqnarray}

For the pure gauge theory, the field equations (\ref{su2fieldeq}) don't 
fix unambigously the components of gluon field $A^a_\mu$. In particular, the
determination of $\theta_1$ and $\theta_2$ requires additional conditions 
on $A^a_\mu$, i.e. one has to rely on a gauge fixing condition.

So far we have considered the field equations associated to the gluon field
built after identifying $n$ with $e_1$. 
Different choices for $n$ will reproduce essentially the picture just 
described. The simplest gluon field is obtained with $n \, = \, e_3$. Then 
\begin{eqnarray}
 & A^1_\mu & = \, \sin \theta_2 \, \hat{A}_\mu \, , \\
 & A^2_\mu & = \, - \cos \theta_2 \, \hat{A}_\mu \, , \\
 & A^3_\mu & = \, \frac{1}{g} \, \partial_\mu \theta_2 \, ,
\end{eqnarray}
the gluon field tensor being
\begin{equation}
  F^a_{\mu \nu} \, = \, n^a \, \mathcal{F}_{\mu \nu} \, ,
\end{equation}
where
\begin{equation}
  \mathcal{F}_{\mu \nu} \, = \, \partial_\mu \hat{A}_\nu \, - \,
                                \partial_\nu \hat{A}_\mu
\end{equation}
and the classical equations of motion of motion are
\begin{equation}
  \partial^\nu \, \mathcal{F}_{\mu \nu} \, = \, 0 \, .
  \label{su2fieldequations}
\end{equation}
For this gluon field the equivalence between classical SU(2) and an abelian
theory is even more striking. The hamiltonian density is a hamiltonian 
density of the ``abelian theory'' associated to $\hat{A}$,
\begin{equation}
   \mathcal{H} \, = \, \mathcal{F}^{\beta 0} \, \partial^0 \hat{A}_\beta \, -
        \, \mathcal{L} 
  \label{su2H}
\end{equation}
and the spin tensor reproduces the spin tensor of the same ``abelian theory''
\begin{equation}
   \mathcal{S}^{\alpha \beta} \, = \, \mathcal{F}^{\beta 0} \,  \hat{A}^\alpha
                 \, - \, \mathcal{F}^{\alpha 0} \,  \hat{A}^\beta \, .
 \label{su2S}
\end{equation}
As previously, the field equations don't fix completely $A^a_\mu$ and to
compute the gluon fields one has to work on a particular gauge. The Landau
gauge condition
\begin{equation}
   \partial^\mu \, A^a_\mu \, = \, 0 \, ,
\end{equation}
now reads
\begin{equation}
  \partial^\mu \, \hat{A}_\mu \, = \, 0 \, ,
  \hspace{1cm}
   \partial^\mu \theta_2 \, \hat{A}_\mu \, = \, 0 \, ,
  \hspace{1cm}
  \partial_\mu \partial^\mu \, \theta_2 \, = \, 0 \, ,
  \label{su2landau}
\end{equation}
i.e. the ansatz reduces the nonlinear field equations to a set of linear
equations (\ref{su2fieldequations}) and a linear gauge condition to a set of 
coupled equations.

Let us discuss the properties of solutions of (\ref{su2fieldequations}) and 
(\ref{su2landau}). Particular solutions of the gauge fixing condition are
\begin{equation}
 \begin{array}{lcl}
     \mbox{Type I}  & \hspace{0.7cm} & \partial_\mu \theta_2 \, = \, 0 \, ,\\
     \mbox{Type II} &  & \hat{A}_\mu \, = \, 0 \, .
 \end{array}
\end{equation}
For type I solutions, the field equations are the Maxwell equations and the
classical finite action SU(2) configurations are the finite action QED-like
configurations  associated with the vector field $\hat{A}_\mu$; $\theta_2$
is a constant field. 
The other family of solutions have null action, energy and spin. They are
interesting because type II solutions includes a new class of configurations. 
The function $\theta_2$ is a solution of the Klein-Gordon equation. The 
requirement of finite action does not constraint $\theta_2$ and the general 
solution is
\begin{eqnarray}
 &  \theta_2 (x) \, = & \int \, \frac{d^3 k}{(2 \pi)^3 \, 2 k_0} \,
 \left[ a( \vec{k} ) \, e^{-i kx} \, + \, 
        a^* ( \vec{k} ) \, e^{ i k x } \right] \, + \nonumber \\
  & &
 \int \, \frac{d^3 k}{(2 \pi)^3 \, 2 k_0} \,
       \left[ b( \vec{k} ) \, e^{kx} \, + \, 
        c( \vec{k} ) \, e^{- k x } \right] \, ,
\end{eqnarray}
where it was assumed that all integrations are well defined.
$\theta_2$ is a linear combination of exponentional functions characterized 
by a four-vector $k$ satisfying the condition $k^2 \, = \, 0$. The components
proportional to $a$ are eigenfunctions of $-i \partial_\mu$ associated to real
eigenvalues $\pm k_\mu$. These components can be identified with free gluons.
The components proportional to $b$ or $c$ are eigenfunctions of the same 
operator but the associated eigenvalues are pure imaginary numbers 
$ \pm i k_\mu$, i.e. they cannot describe free particles. 
Considering that free gluons have never been observed in nature, 
configurations of the last kind should not be disregarded \textit{a priori}. 
Actually, the same observation may suggest that the components 
that one should disregard are those proportional to $a$. 
The above reasoning is valid for a linear theory like QED. For SU(2), the 
identification of free field components with eigenfunctions of 
$-i \partial_\mu$ is arguable and this naive reasoning should be taken with
care.

The general solution of the field equations with finite action in the Landau 
gauge is
\begin{eqnarray}
  \hat{A}_\mu  & = & \sum\limits^{}_{\lambda} \, a( \vec{k} , \lambda ) \,
\epsilon_\mu ( k , \lambda ) \, e^{ \pm i kx} \, ,
 \label{su2gensolA} \\
   \theta_2    & = & \left[ a( \vec{k} ) \, e^{-i kx} \, + \, 
                      a^* ( \vec{k} ) \, e^{ i k x } \right] \, + \nonumber \\
               &   & \hspace{1.5cm}
                     \left[ b( \vec{k} ) \, e^{kx} \, + \, 
                            c( \vec{k} ) \, e^{- k x } \right] \, ,
 \label{su2gensolt}
\end{eqnarray}
where $k^2 \, = \, 0$ and the three independent polarization vectors 
$\epsilon_\mu ( k , \lambda )$ verify the conditions
\begin{equation}
   k^\mu \epsilon_\mu ( k , \lambda ) \, = \, 0 \, ,
  \qquad \lambda \, = \, 1 \, \dots \, 3 \qquad .
\end{equation}
For the special case of $k \, = \, ( \omega, \, 0, \, 0, \, \omega)$, possible
choices for the polarization vectors are
\begin{equation}
\epsilon_\mu ( k ,1 ) \, = \, k_\mu \, / \, \omega \, , \qquad
\epsilon_\mu ( k ,2 ) \, = \, \left( \begin{array}{c}
                                        0 \\
                                        1 \\
                                        0 \\
                                        0
                                     \end{array} 
                               \right) \, , \qquad
\epsilon_\mu ( k ,3 ) \, = \, \left( \begin{array}{c}
                                        0 \\
                                        1 \\
                                        0 \\
                                        0
                                     \end{array} 
                               \right)\, .
\end{equation}
However, similarly to classical electrodynamics, one can redefine the vector
field $\hat{A}_\mu$ by adding up a gradient of a scalar function $\eta$ such
that $\partial^2 \eta \, = \, 0$ and 
$\partial_\mu \theta_2 \, \partial^\mu \eta \, = \, 0$. The particular choice
$\eta \, = \, \pm i \, a(k,1) \, e^{\pm ikx} \, / \, \omega$
satisfies the above requirements and removes the longitudinal component from 
$\hat{A}_\mu$. The vector field $\hat{A}_\mu$ can be made transverse but the
gluon field built from (\ref{su2gensolA}) - (\ref{su2gensolt}) has a
longitudinal component associated to $\theta_2$. 
Remember that energy and spin are independent of the longitudinal component of
the gluon field and that $\theta_2$ plays a role in the interaction with matter
fields. In what concerns only pure gauge SU(2) theory, the finite
action classical solutions are essentially QED-like solutions associated to
$\hat{A}_\mu$ field.

A special class of configurations are the classical vacuum solutions. They
are solutions with a null gluon field tensor,
\begin{equation}
 F^a_{\mu \nu} \, = \, 0 \, .
\end{equation}
For the ansatz considered, this means
\begin{equation}
  \hat{A}_\mu \, = \, \partial_\mu \chi \, ,
\end{equation}
where $\chi$ is any differentiable function of spacetime. Vaccum 
configurations are determined by two scalar functions $\chi$ and $\theta_2$.
The field equations don't give us information about the nature of these
functions. The Landau gauge condition requires that
\begin{equation}
  \partial_\mu \partial^\mu \, \chi \, = \, 0 \, ,
  \hspace{1cm}
   \partial^\mu \theta_2 \, \partial_\mu  \chi\, = \, 0 \, ,
  \hspace{1cm}
  \partial_\mu \partial^\mu \, \theta_2 \, = \, 0 \, ,
  \label{su2vaclandau}
\end{equation}
whose general solution is, again, characterized by a four-vector $k$
satisfying $k^2 \, = \, 0$,
\begin{eqnarray}
   \chi    & = & \left[ a_\chi( \vec{k} ) \, e^{-i kx} \, + \, 
                        a^*_\chi ( \vec{k} ) \, e^{ i k x } \right] \, + 
                 \nonumber \\
           &   & \hspace{1.5cm}
                 \left[ b_\chi( \vec{k} ) \, e^{kx} \, + \, 
                        c_\chi( \vec{k} ) \, e^{- k x } \right] \, , 
  \label{vacchi} \\
   \theta_2    & = & \left[ a_\theta ( \vec{k} ) \, e^{-i kx} \, + \, 
                            a^*_\theta ( \vec{k} ) \, e^{ i k x } \right] \, + 
                 \nonumber \\
               &   & \hspace{1.5cm}
                     \left[ b_\theta( \vec{k} ) \, e^{kx} \, + \, 
                            c_\theta( \vec{k} ) \, e^{- k x } \right] \, .
  \label{vactheta}
\end{eqnarray}
Solutions (\ref{vacchi}) and (\ref{vactheta}) are similar to the finite action
solutions discussed previously. The main difference being that now the
gluon field is pure longitudinal and, as before, $\chi$ can be chosen such that
$\hat{A}_\mu \, = \, 0$. Then, the vacuum field is
\begin{equation}
 A^a_\mu \, = \, \delta^{a3} \, k_\mu \, \left\{ ~
   \left[ a ( \vec{k} ) \, e^{-i kx} \, + \, 
                            a^* ( \vec{k} ) \, e^{ i k x } \right] \, + 
                     \left[ b( \vec{k} ) \, e^{kx} \, + \, 
                            c( \vec{k} ) \, e^{- k x } \right] \, ~ \right\}.
\end{equation}

Our study of SU(2) is a first flavor for the classical configurations of
SU(3) pure gauge theory. As we will see in the next section, the 
``abelian projection'' and the main characteristics of the SU(2) solutions
are also present in the SU(3) classical configurations.

\section{Classical SU(3) Gauge Theory}

To solve the classical equations of motion for QCD, we choose a
spherical like basis in the color space - see appendix for definitions.
Setting $n \, = \, \vec{e}_3$, condition (\ref{Dn}) becomes
\begin{equation}
 - \frac{1}{2} \partial_\mu n^a \, - \, g f_{abc} \, Y^b_\mu \, n^c \, = \, 0
 \, .
\label{Dn2}
\end{equation}
This set of equations provides the following relations between the $Y^a_\mu$ 
fields
\begin{eqnarray}
 & &  Y^2_\mu \, = \, - \, Y^1_\mu \, \cot \theta_2 \, , \label{Dn01}  \\
 & &  Y^3_\mu \, = \, - \, \frac{1}{2g} \, \partial_\mu \theta_2 \, ,  \\
 & &  Y^4_\mu \, = Y^5_\mu \, = \, Y^6_\mu \, = \, Y^7_\mu \, = \, 0 \, . 
\label{Dn02}
\end{eqnarray}
Taking into account (\ref{Dn01}) to (\ref{Dn02}) the gluon field is given by
\begin{equation}
 \left( A^a_\mu \right) \, = \left( \begin{array}{c}
          - \sin \theta_2 \, \hat{A}_\mu \, + \, Y^1_\mu \\
            \cos \theta_2 \, \hat{A}_\mu \, - \, \cot \theta_2 \, Y^1_\mu \\
            \partial_\mu \theta_2 \, / \, g  \\
            0 \\
            0 \\
            0 \\
            0 \\
            Y^8_\mu
                                    \end{array} \right) \, .
\end{equation}
Condition (\ref{nY}) simplifies the gluon field into
\begin{equation}
 A^a_\mu \, = \, n^a \hat{A}_\mu \, + \, 
                 \delta^{a3} \, \frac{1}{g} \, \partial_\mu \theta_2
                 \, + \, \delta^{^a8} C_\mu \, ;
 \label{glue}
\end{equation}
in the last equation $C_\mu \, = \, Y^8_\mu$.
The corresponding gluon field tensor components are
\begin{equation}
  F^a_{\mu \nu} \, = \, n^a \, \mathcal{A}_{\mu\nu} \, + \,
           \delta^{a8} \, \mathcal{C}_{\mu \nu}
\end{equation}
where
\begin{equation}
 \mathcal{A}_{\mu \nu} \, = \, \partial_\mu \hat{A}_\nu \, - \,
                               \partial_\nu \hat{A}_\mu \, , \hspace{0.7cm}
 \mathcal{C}_{\mu \nu} \, = \, \partial_\mu C_\nu \, - \,
                               \partial_\nu C_\mu \, .
\end{equation}
The classical action is a functional of the ``photon like'' fields
$\hat{A}_\mu$ and $C_\mu$,
\begin{equation}
  \mathcal{L} \, = \, - \frac{1}{4} \left(
    \mathcal{A}^2 \, + \, \mathcal{C}^2 \right) \, ,
\end{equation}
and the classical equations of motion
\begin{equation}
 D^\nu F^a_{\mu \nu} \, = \, 
    \partial^\nu F^a_{\mu \nu} \, - \,
    g f_{abc} A^b_\nu F^c_{\mu \nu} \, = \, 0
\end{equation}
are reduced to
\begin{eqnarray}
 \partial_\nu \mathcal{A}^{\mu \nu} \, = \, 0 \, ,
 & &
 \partial_\nu \mathcal{C}^{\mu \nu} \, = \, 0 \, .
 \label{equacoes}
\end{eqnarray}
The hamiltonian density
\begin{equation}
 \mathcal{H} \, = \, \mathcal{A}^{\beta 0} \, \partial^0 \hat{A}_\beta \, + \,
                     \mathcal{C}^{\beta 0} \, \partial^0 C_\beta \, - \,
                     \mathcal{L}
\end{equation}
and the spin tensor
\begin{equation}
 \mathcal{S} \, = \, \left( \mathcal{A}^{\beta 0} \, \hat{A}^\alpha \, - \,
                            \mathcal{A}^{\alpha 0} \, \hat{A}^\beta \right) 
    \, + \,
                      \left( \frac{}{} 
                             \mathcal{C}^{\beta 0} \, C^\alpha \, - \,
                            \mathcal{C}^{\alpha 0} \, C^\beta \right)
\end{equation}
are given by the sum of the contributions of two abelian-like theories
associated with $\hat{A}$ and $C$. Similarly to what was observed for the 
SU(2) theory, the ansatz (\ref{glue}) maps SU(3) to a set of two linear 
theories, making classical pure SU(3) gauge theory formally equivalent to QED
with two ``photon fields''. Note that the  ``photon fields'' are not coupled 
and that their coupling to the fermionic fields requires different
Gell-Mann matrices.

The classical equations of motion (\ref{equacoes}) are independent of
$\theta_2$. Therefore, to compute classical configurations we consider the
Landau gauge. In terms of $\hat{A}_\mu$, $C_\mu$ and $\theta_2$ the gauge
condition reads
\begin{eqnarray}
 \partial^\mu \hat{A}_\mu \, = \, 0, &  & 
             \partial^\mu C_\mu \, = \, 0  \, , \nonumber \\
 (\partial^\mu \theta_2) \hat{A}_\mu \, = \, 0  \, , & &
              \Box \theta_2 \, = \, 0 \, .
 \label{landau}
\end{eqnarray}
The solutions of (\ref{equacoes}) and (\ref{landau}) are similar to the
classical configurations discussed for SU(2). The main difference comes from
having now two ``photon'' fields instead of a single one. The general finite
action solution is then given by
\begin{eqnarray}
  \hat{A}_\mu  & = & \sum\limits^{}_{\lambda} \, a_A( \vec{k} , \lambda ) \,
\epsilon_\mu ( k , \lambda ) \, e^{ \pm i kx} \, ,
 \label{su3gensolA} \\
   \theta_2    & = & \left[ a( \vec{k} ) \, e^{-i kx} \, + \, 
                      a^* ( \vec{k} ) \, e^{ i k x } \right] \, + \nonumber \\
               &   & \hspace{1.5cm}
                     \left[ b( \vec{k} ) \, e^{kx} \, + \, 
                            c( \vec{k} ) \, e^{- k x } \right] \, ,
 \label{su3gensolt} \\
  C_\mu  & = & \sum\limits^{}_{\lambda} \, a_C( \vec{k} , \lambda ) \,
\epsilon_\mu ( k , \lambda ) \, e^{ \pm i kx} \, ,
 \label{su3gensolC}
\end{eqnarray}
where $k^2 \, = \, 0$ and the sum over polarizationas runs only over the
transverse polarizations.

Let us discuss the classical vacuum solutions of (\ref{equacoes}) and 
(\ref{landau}). Following the approach as for SU(2), one can set
$\hat{A}_\mu \, = \, C_\mu \, = \, 0$, i.e. 
\begin{equation}
  A^a_\mu \, = \, \delta^{a3} \, \frac{1}{g} \, \partial_\mu \theta_2 \, .
\end{equation}
Then, (\ref{landau}) reduces to
\begin{equation}
   \Box \theta_2 \, = \, 0 \, 
 \label{teta}
\end{equation}
and that $\theta_2$ does not verify any special kind of boundary
condition. The above equation can be solved by separation of variables in the
usual way. In spherical coordinates, the solution is
\begin{eqnarray}
    \theta_2 \, = \, \sum\limits^{+ \infty}_{l = 0} \, 
                     \sum\limits^{l}_{m = -l} \, 
    & &
      \left\{ \frac{ }{ } \right.
  \left[ \frac{}{} a (\omega) e^{- i \omega t} \, + \,
         a^* (\omega) e^{+ i \omega t}  \right] \,  \times
  \nonumber \\
  & & \hspace{1.2cm}
     \left[ \frac{}{} A_{lm} ( \omega ) j_l ( \omega r ) \, + \,
            B_{lm} ( \omega ) n_l ( \omega r ) \right] \, 
        Y_{lm} ( \theta , \phi )
     \nonumber \\
 & &  + \,
  \left[ \frac{}{}  b (\omega) e^{-  \omega t} \, + \,
         c (\omega) e^{+  \omega t}  \right] \,\times
  \nonumber \\
  & & \hspace{1.2cm}
     \left[ \frac{}{} C_{lm} ( \omega ) j_l ( i \omega r ) \, + \,
            D_{lm} ( \omega ) n_l ( i \omega r ) \right] \,
      Y_{lm} ( \theta , \phi )
 \nonumber \\
 & & + \,
  \left[ \frac{}{} e t \, + \, f \right] 
     \left[ F_{lm} r^l \, + \,
            \frac{G_{lm}}{r^{l+1}} \right]
      Y_{lm} ( \theta , \phi )
   \left. \frac{}{} \right\} \, , \nonumber \\
 & &
 \label{theta}
\end{eqnarray}
where $\omega$ has dimensions of mass, $j_l$ is the spherical Bessel function
of order $l$ and $n_l$ is the spherical Neumann function of order $l$. The
parameter $\omega$ is related to the four-vector $k$ 
($k_0 \, = \, | \vec{k} | \, = \, \omega$) and cames into (\ref{theta}) when
the separation between time and spacial parts of $\theta_2$ is performed.
Note that, as in SU(2), the classical gluon field vacuum is a pure 
longitudinal field.

\section{Quarks in the Background of Classical Configurations}

Let us discuss now the properties of quark fields propagating in the
background of the classical SU(3) configurations computed in the previous
section. In terms of color components, the Dirac equation
\begin{equation}
 \left( i D \!\!\!\! / ~ - ~ m \right) \, \psi \, = \, 0
\end{equation}
reads
\begin{eqnarray}
 & & ( i \partial \!\!\!\! /  \, - \, m ) \, \psi^1 ~ + ~
     i \, \frac{g}{2} \, e^{- i \theta_2 } \, \hat{A}\!\!\!\! / \, \psi^2 ~ - ~
     \frac{1}{2} \, \partial \!\!\!\! / \theta_2 \, \psi^1 ~ - ~
     \frac{g}{2 \sqrt{3}} C \!\!\!\! / \, \psi^1 ~ = ~ 0 \, ,
 \label{dirac1} \\
 & & ( i \partial \!\!\!\! /  \, - \, m ) \, \psi^2 ~ - ~
     i \, \frac{g}{2} \, e^{ i \theta_2 } \, \hat{A}\!\!\!\! / \, \psi^1 ~ + ~
     \frac{1}{2} \, \partial \!\!\!\! / \theta_2 \, \psi^2 ~ - ~
     \frac{g}{2 \sqrt{3}} C \!\!\!\! / \, \psi^2 ~ = ~ 0  \, ,
\label{dirac2} \\
 & & ( i \partial \!\!\!\! /  \, - \, m ) \, \psi^3 ~ + ~
     \frac{g}{\sqrt{3}} C \!\!\!\! / \, \psi^3 ~ = ~ 0  \label{dirac3} \, .
\end{eqnarray}
Introducing the spinor fields $\phi$ and $\eta$ defined by
\begin{equation}
  \psi^1 ~ = ~ e^{ - \frac{i}{2} \, \theta_2 } ~ \phi \, , 
  \hspace{1cm}
  \psi^2 ~ = ~ i ~ e^{ \frac{i}{2} \, \theta_2 } ~ \eta \, ,   
\end{equation}
(\ref{dirac1})-(\ref{dirac3}) can be writen as a set of uncoupled equations
\begin{eqnarray}
 & & \left[ i \partial \!\!\!\! /  ~ - ~ m  ~ - ~
     \frac{g}{2} \left( \frac{C \!\!\!\! /}{\sqrt{3}} \, - \,
                        \hat{A} \!\!\!\! / \right) \right] \phi^{(+)} ~ = ~ 0
 \, ,
 \label{ddirac1} \\
 & & \left[ i \partial \!\!\!\! /  ~ - ~ m  ~ - ~
     \frac{g}{2} \left( \frac{C \!\!\!\! /}{\sqrt{3}} \, + \,
                        \hat{A} \!\!\!\! / \right) \right] \phi^{(-)} ~ = ~ 0
  \, ,\label{ddirac2} \\
 & & ( i \partial \!\!\!\! /  \, - \, m ) \, \psi^3 ~ + ~
     \frac{g}{\sqrt{3}} C \!\!\!\! / \, \psi^3 ~ = ~ 0  \label{ddirac3} \, ,
\end{eqnarray}
where
\begin{equation}
   \phi^{( \pm )} ~ = ~ \phi \, \pm \, \eta \, .
\end{equation} 
The vector fields $\hat{A}_\mu$ and $C_\mu$ are electromagnetic-type fields
and the problem of solving the Dirac equation is reduced to a set of three
electromagnetic-like problems. Note that in (\ref{dirac1})-(\ref{dirac3}) the
electromagnetic couplings cannot all be atractive couplings. Indeed, the 
maximum number of attractive couplings is two. In terms of dynamics
this means that, at least, for one the equations the lowest energy solution
is the free particle solution of mass $m$ at rest.

Let $\{ \phi^{(+)}_n, \, E^{(+)}_n \}$, $\{ \phi^{(-)}_n, \, E^{(-)}_n \}$,
$\{ \phi^{(3)}_n, \, E^{(3)}_n \}$ be the Dirac spinor and associated 
energy for the solutions of equations (\ref{ddirac1}), (\ref{ddirac2}), 
(\ref{ddirac3}) respectively. Then, the original Dirac field is given by
\begin{eqnarray}
& & \psi^1 (x) \, = \, \frac{e^{-\frac{i}{2} \theta_2}}{2} ~
 \sum\limits^{}_{n} \, \left[  ~  \phi^{(+)}_n ( \vec{x} ) \,\, 
                                                e^{- i E^{(+)}_n t}
                                  \, + \,
                                  \phi^{(-)}_n ( \vec{x} ) \,\,
                                                e^{- i E^{(-)}_n t} ~
                        \right] \, ,
  \label{cor1} \\
  & & \psi^2 (x) \, = \, i ~ \frac{e^{\frac{i}{2} \theta_2}}{2} ~
 \sum\limits^{}_{n} \, \left[  ~  \phi^{(+)}_n ( \vec{x} ) \,\,
                                                e^{-i E^{(+)}_n t}  \, - \,
                                  \phi^{(-)}_n ( \vec{x} ) \,\,
                                                e^{-i E^{(-)}_n t} ~
                        \right] \, ,
  \label{cor2} \\
  &  & \psi^3 (x) \, = \, \sum\limits^{}_{n} \left[ ~
                                         \phi^{(3)}_n ( \vec{x} ) \,\,
                                         e^{-i E^{(3)}_n t}  
                                                ~   \right] \, .
  \label{cor3}
\end{eqnarray}
If one identifyies physical states with eigenstates of the operator
$i \partial_0$, the candidates to be physical particles are
\begin{eqnarray}
  \mbox{any product of } \, \psi^3 \, , & &
  \mbox{products of } \, \psi^2 \, \psi^1\, , \nonumber \\
  \sum\limits^{}_a {\overline \psi}^a \, \Gamma \, \psi^a \, , &  &
  \epsilon_{abc} ~ \chi_{\alpha\beta\gamma} ~
                      \psi^a_\alpha \, \psi^b_\beta \, \psi^c_\gamma \, ;
  \label{particulas0}
\end{eqnarray}
roman letters stand for color indices and greek letters for spin indices. In
(\ref{particulas0}), the first and second wave functions are not gauge
invariant, therefore the associated eigenvalue of $i \partial_0$ is not gauge
invariant\footnote{by a gauge transformation the wave function can be made
time independent. A state with zero $i \partial_0$ eigenvalue.} and they cannot
be associated to physical particles. On the other hand, the second and third
types of wave functions are gauge invariant and their energy spectrum 
can be identified with physical states. These wave functions are the usual
quark model wave functions for mesons and baryons. 

Let us assume that physical particles are described by the following wave
functions
\begin{equation}
  M \, = \, 
     \sum\limits^{}_a {\overline \psi}^a \, \Gamma \, \psi^a \, , \hspace{1cm}
  B \, = \, \epsilon_{abc} ~ \chi_{\alpha\beta\gamma} ~
                      \psi^a_\alpha \, \psi^b_\beta \, \psi^c_\gamma \, .
  \label{particulas}
\end{equation}
Taking into account (\ref{cor1})-(\ref{cor3}), the meson wave functions
are given by
\begin{eqnarray}
 &  M (x) \, = \, 
  \frac{1}{2} \sum\limits^{}_{n,k} 
  \Big\{  & \left[ {\overline \phi}^{(+)}_n  (\vec{x}) ~ \Gamma ~ 
                              \phi^{(+)}_k (\vec{x})  \right] \,
            e^{-i \left( E^{(+)}_k - E^{(+)}_n \right) t} \, + \, 
 \nonumber \\
 & &
\left[ {\overline \phi}^{(-)}_n  (\vec{x}) ~ \Gamma  ~
                   \phi^{(-)}_k (\vec{x}) \right] \,
            e^{-i \left( E^{(-)}_k - E^{(-)}_n \right) t} \, + \,
 \nonumber \\
 & &
\left[ {\overline \phi}^{(3)}_n  (\vec{x})  ~ \Gamma ~
                  \phi^{(3)}_k (\vec{x}) \right] \,
            e^{-i \left( E^{(3)}_k - E^{(3)}_n \right) t} ~~
 \Big\} \label{mesao}
\end{eqnarray}
while the baryon wave functions are
\begin{eqnarray}
 &  B \, \sim \, 
  \sum\limits^{}_{n_1,n_2,n_3} 
  \Big\{  & A_{n_1 \, n_2 \, n_3} \, ~
      e^{-i \left( E^{(+)}_{n_1} +  E^{(+)}_{n_2} + E^{(3)}_{n_3}  \right) t} 
       \, + \, 
 \nonumber \\
 & &   B_{n_1 \, n_2 \, n_3} \, ~
      e^{-i \left( E^{(+)}_{n_1} +  E^{(-)}_{n_2} + E^{(3)}_{n_3}  \right) t} 
       \, + \,
 \nonumber \\
 & &  C_{n_1 \, n_2 \, n_3} \, ~
      e^{-i \left( E^{(-)}_{n_1} +  E^{(-)}_{n_2} + E^{(3)}_{n_3}  \right) t} 
 \Big\} \, , \label{bariao}
\end{eqnarray}
where $A_{n_1 \, n_2 \, n_3}$, $B_{n_1 \, n_2 \, n_3}$ and 
$C_{n_1 \, n_2 \, n_3}$ are the baryon spatial wave function. In this picture,
mesons and baryons are described as systems of non-interacting quarks in
the background of the classical configurations considered in last section. Of 
course, this is not the true picture of a meson or a baryon but 
a first order approximation to hadrons. This way of viewing hadronic matter
is identical to the description of the atomic and nuclear energy levels.

The energy spectrum described by $E^{(+)}_n$, $E^{(-)}_n$ and $E^{(3)}_n$ are
electromagnetic spectra. Assuming that the quark-gluon interaction can be 
viewed as static interaction, described by a Coulombic potential, one can
immediatly compute the associated particles masses. Before starting to compare
the theoretical spectrum with measured masses, one should first try to
identify the family of hadrons that fits better such a theoretical 
description. The success of heavy quark effective field theory 
\cite{Neu94,KP94} and nonrelativistic potential studies of heavy
mesons \cite{LP96} suggests that heavy mesons are good testing grounds
for the above hypotheses. This idea is corroborated by quenched lattice QCD
investigations, an approximation that solves exactly the bosonic sector of
the theory and propagates quarks in the gluon fields. It is well known that
such an approximation provides better results for heavy
quark systems \cite{MM94}. Furthermore, it is well known that charmonium and
bottomonium spectrum are close to hidrogen-like spectrum. 
All this observations motivates the next section.

\section{Classical Gluonic Configurations, \\ Charmonium and Bottomonium 
Spectrum}

According to our picture of hadrons, mesons and baryons are classified with
hidrogen-like levels. The main difference to the hidrogen atom being that now
there are two independent sequence of levels, i.e. the Coulombic potential is
not an interquark potential but describes the interaction between a quark and
a background gluonic field. According to (\ref{mesao}),
the two independent spectrum do not mix.

The solution of the Dirac equation in a Coulomb potential is well known
\cite{IZ80}. The energy spectrum is given by
\begin{equation}
 E_{n ~ j} \, = \, \frac{m}
                     {\sqrt{1 + \frac{\alpha^2}
                                     {\left( n - \delta_j \right)^2}}}
 \, , \hspace{1cm} \delta_j \, = \, j \, + \, \frac{1}{2} \, - \,
  \sqrt{ \left( j \, + \, \frac{1}{2} \right)^2 - \alpha^2 } \, ,
 \label{coulomb}
\end{equation}
with $n \, = \, 1, \, 2, \, \dots$ and 
$j \, = \, \frac{1}{2}, \, \frac{3}{2}, \, \dots , \, n - \frac{1}{2}$.
The spectrum has a twofold degeneracy except for the state 
$j \, = \, n - \frac{1}{2}$. This degenerate states can be distinguished by
their orbital angular momentum $l \, = \, j \, \pm  \frac{1}{2}$ (for
$j \, = \, n - \frac{1}{2}$, $l \, = \, n - 1$). The quarkonium levels 
are given by (\ref{coulomb}) with the coupling constant $\alpha$ replaced
 by the strong coupling constant multiplied by a factor $F^2$, 
\begin{equation}
  F \, = \, \Bigg\{
      \frac{1}{2} \left( a \, - \, \frac{c}{\sqrt{3}} \right) , ~
   - ~ \frac{1}{2} \left( a \, + \, \frac{c}{\sqrt{3}} \right) , ~
       \frac{c}{\sqrt{3}}
            \Bigg\} \, ,
\end{equation}
where $a$ and $c$ are, respectively,
the $\hat{A}_\mu$ and $C_\mu$ amplitudes; see 
(\ref{ddirac1})-(\ref{ddirac3}). In the electromagnetic spectrum, the
Lamb shift resolves the degeneracy between $n \, s_j$ and $n \, p_j$ states, 
the $s$ states acquire an excess of energy. In our calculation of the
particles masses we will not take into account such a correction but,
certainly, a more precise computation should take into account the Lamb
shift and corrections to the particle-independent approximation assumed
in our description of hadrons.

As quark masses we take the central values of the 2002 edition of the Particle
Data Book \cite{pdg02},
\begin{equation}
 m_c \, = \, 1200 ~ \mbox{MeV} \, , \hspace{1cm}
 m_b \, = \, 4850 ~ \mbox{MeV} \, .
 \label{massas}
\end{equation}
For the strong coupling constants we use the central values given in
\cite{alphaS},
\begin{equation}
 \alpha_s (m_c) \, = \, 0.42124 \,  , \hspace{1cm}
 \alpha_S (m_b) \, = \, 0.21174  \, .
 \label{alfa}
\end{equation}
In the following it is assumed that the quantum numbers of charmonium and
bottomonium given in the Particle Data Book are correct. Table \ref{bbcc}
is a resum\'e of particles masses and quantum numbers.

\begin{table}
\begin{center}
\begin{tabular}{cc}
\begin{tabular}{||l||@{\hspace{0.3cm}}c@{\hspace{0.3cm}}|@{\hspace{0.3cm} }c@{\hspace{0.3cm}}||}
 \hline
 \hline
                 &  $J^P$  &  M \\
 \hline
 \hline
  $\eta_b (1S)$  &  $0^-$  & $9300 \pm 20 \pm 20$ \\
 \hline
  $\Upsilon_b (1S)$  &  $1^-$  & $9460.30 \pm 0.26$ \\
 \hline
  $\chi_{b0} (1P)$  &  $0^+$  & $9859.9 \pm 1.0$ \\
  $\chi_{b1} (1P)$  &  $1^+$  & $9892.7 \pm 0.6$ \\
  $\chi_{b2} (1P)$  &  $2^+$  & $9912.6 \pm 0.5$ \\
\hline
  $\Upsilon_b (2S)$  &  $1^-$  & $10023.26 \pm 0.31$ \\
\hline
  $\chi_{b0} (2P)$  &  $0^+$  & $10232.1 \pm 0.6$ \\
  $\chi_{b1} (2P)$  &  $1^+$  & $10255.2 \pm 0.5$ \\
  $\chi_{b2} (2P)$  &  $2^+$  & $10268.5 \pm 0.4$ \\
\hline
  $\Upsilon_b (3S)$  &  $1^-$  & $10355.2 \pm 0.6$ \\
\hline
  $\Upsilon_b (4S)$  &  $1^-$  & $10580.0 \pm 3.5$ \\
\hline
  $\Upsilon_b (10860)$  &  $1^-$  & $10865 \pm 8$ \\
\hline
  $\Upsilon_b (11020)$  &  $1^-$  & $11019 \pm 8$ \\
\hline
\hline
\end{tabular} &
\begin{tabular}{||l||@{\hspace{0.3cm}}c@{\hspace{0.3cm}}|@{\hspace{0.3cm} }c@{\hspace{0.3cm}}||}
 \hline
 \hline
                 &  $J^P$  &  M \\
 \hline
 \hline
  $\eta_c (1S)$  &  $0^-$  & $2979.7 \pm 1.5$ \\
 \hline
  $J/ \psi (1S)$  &  $1^-$  & $3096.87 \pm 0.04$ \\
 \hline
  $\chi_{c0} (1P)$  &  $0^+$  & $3415.1 \pm 0.8$ \\
  $\chi_{c1} (1P)$  &  $1^+$  & $3510.51 \pm 0.12$ \\
  $\chi_{c2} (1P)$  &  $2^+$  & $3556.18 \pm 0.13$ \\
\hline
  $h_c (1P)$  &  $?^?$  & $3526.14 \pm 0.24$ \\
\hline
  $\eta_c (2S)$  &  $0^-$  & $3594 \pm 5$ \\
\hline
  $\psi (2S)$  &  $1^-$  & $3685.96 \pm 0.09$ \\
\hline
  $\psi (3770)$  &  $1^-$  & $3769.9 \pm 2.5$ \\
\hline
  $\psi (3836)$  &  $2^-$  & $3836 \pm 13$ \\
\hline
  $\psi (4040)$  &  $1^-$  & $4040 \pm 10$ \\
\hline
  $\psi (4160)$  &  $1^-$  & $4159 \pm 20$ \\
\hline
  $\psi (4415)$  &  $1^-$  & $4415 \pm 6$ \\
\hline
\hline
\end{tabular}
\end{tabular}
\caption{Bottomonium anc charmonium spectrum according to
Particle Data Book \cite{pdg02}.
\label{bbcc}}
\end{center}
\end{table}

For the double hidrogenium spectrum, the sequence of the lowest energy levels
is
\begin{equation}
 1s^2_{\frac{1}{2}} (0^-), ~
 \left( 1s_{\frac{1}{2}}, 2p_{\frac{1}{2}} \right) (0^+, \, 1^+), ~
 \left( 1s_{\frac{1}{2}}, 2s_{\frac{1}{2}} \right) (0^-, \, 1^-), ~
 \left( 1s_{\frac{1}{2}}, 2p_{\frac{3}{2}} \right) (1^+, \, 2^+);
 \label{primeiros}
\end{equation}
states are identified with the usual spectroscopic notation and
in parentheses are writen the allowed $J^P$ values. 
In each of the sequences in table \ref{bbcc}, the first quarkonium
states can be interpreted as same of the (\ref{primeiros}) states. Note that,
for charmonium $\chi (1P)$ states, the difference in mass between the $0^+$ 
and $1^+$ is almost twice the difference between the $1^+$ and $2^+$ 
states. Then, one can take as the mass of the 
$\left( 1s_{1/2}, 2p_{3/2} \right) (1^+, \, 2^+)$ state, the
mean value of $\chi_{c1}$ and $\chi_{c2}$ masses and identify
$\chi_{c0}$ as 
$\left( 1s_{1/2}, 2p_{1/2} \right) (0^+)$. 
Then, the mass difference between these two states is given by
\begin{equation}
 E_{2 ~ 3/2} ~ - ~ E_{2 ~ 1/2} \, = \, \Delta M (F^2) \, .
  \label{diferenca}
\end{equation}
From (\ref{diferenca}) one can compute $F$. A similar reasoning applies to the
bottomonium\footnote{The observation also applies to $\chi_b (2P)$ states.}
$\chi_b (1P)$ particles. 
The $F$ values computed using table \ref{bbcc} masses are
\begin{equation}
   F \, = \, \left\{ \begin{array}{lcl}
                      1.5259131 & & \mbox{from } \chi_c (1P), \\
                      1.7781466 & & \mbox{from } \chi_b (1P).
                     \end{array}
             \right.
  \label{F}
\end{equation}
The corresponding effective strong coupling constant being
\begin{equation}
   \alpha (m_c) 
\, = \, \alpha_s (m_c) \, F^2 \, = \, \left\{ \begin{array}{lcl}
                      0.98082 & & \mbox{using the } \chi_c (1P) \mbox{ result},
 \\
                      1.33188 & & \mbox{using the } \chi_b (1P) \mbox{ result},
                                                 \end{array} \right.
  \label{alfac}
\end{equation}
and
\begin{equation}
   \alpha (m_b) 
 \, = \, \alpha_s (m_b) \, F^2 \, = \, \left\{ \begin{array}{lcl}
                      0.49302 & & \mbox{using the } \chi_c (1P) \mbox{ result},
 \\                   0.66948 & & \mbox{using the } \chi_b (1P) \mbox{ result}.
                     \end{array}
             \right.
  \label{alfab}
\end{equation}
Note that one of the $\alpha (m_c)$ values is larger than one. This implies
that there states whose coulombic energy (\ref{coulomb}) is imaginary, meaning
that they are not stable states. Then, according to our description, the 
number of available charmonium states is lower than the number of bottomonium 
states.

In order to reproduce quantitatively the particles masses, one has to compute
the contribution from the gluonic field. A value for the gluonic energy, 
$E_{glue}$, can be obtain using the $F$ values (\ref{F}) and assuming that
\begin{equation}
 M \Big[  \left( 1s_{\frac{1}{2}}, 2p_{\frac{1}{2}} \right) (0^+) \Big]
~ = ~ E_{glue} \, + \, E_{1 ~ \frac{1}{2}} \, + \,
                       E_{2 ~ \frac{1}{2}} ~ .
\end{equation}
Then, using the values of table \ref{bbcc}
\begin{equation}
   E_{glue} \, = \, \left\{ \begin{array}{lcl}
                      2253.66 \mbox{ MeV} & & \mbox{from } \chi_c (1P), \\
                      1729.72 \mbox{ MeV} & & \mbox{from } \chi_b (1P).
                     \end{array}
             \right.
  \label{eglue}
\end{equation}
The gluonic energy (\ref{eglue}) provides a large fraction of the particle
masses. We do not have a justification for the above values but it is 
surprising that the values reported in (\ref{eglue}) are compatible
with quenched lattice QCD estimates for the ground and first excited
states of $J^P \, = \, 0^+$ glueballs masses \cite{Mi03}. In this paper we will
not discuss a possible origin of $E_{glue}$.

We are now in position of discussing the theoretical meson spectrum and
compare it with the experimental data.
For bottomonium the lightest particles are $1 s^2_{1/2} (0^-)$ states. Their
mass values being
\begin{equation}
   M \left[ {\overline b}b; ~ 1 s^2_{1/2} (0^-) \right] \, = \, 
            \left\{ \begin{array}{rcl}
                      8935.20 \mbox{ MeV} & & \mbox{from } \chi_c (1P), \\
                     10692.85 \mbox{ MeV} & & \mbox{from } \chi_b (1P).
                     \end{array}
             \right.
  \label{bb0}
\end{equation}
For charmonium the spectrum associated to $\chi_c (1P)$ has
$1 s^2_{1/2} (0^-)$,
\begin{equation}
   M \left[ {\overline c}c; ~ 1 s^2_{\frac{1}{2}} (0^-) \right] \, = \, 
                      2721.46 \mbox{ MeV}  \hspace{1cm}
                       \mbox{from } \chi_c (1P) \, ,
  \label{cc0c}
\end{equation}
has the lowest energy eigenstate. The stable lowest energy eigenstate
associated to the spectrum computed using $\chi_b (1P)$ data is a
$2 p_{3/2}$ configuration. The lightest meson mass associated to this spectrum
is
\begin{equation}
   M \left[ {\overline c}c; ~ 2 p^2_{\frac{3}{2}} (0^-, \, 2^-) \right] 
   \, = \,  3520.13 \mbox{ MeV}  \hspace{1cm} \mbox{from } \chi_b (1P) \, .
  \label{cc0b}
\end{equation}
According to these results, the lightest quarkonium meson should be a 
$J^P \, = \, 0^-$ meson. This result agrees with the measured data.

For $J^P \, = \, 0^-$ particles our predictions are resumed
in the following tables
\begin{center}
\begin{tabular}[h]{||cl|rl||cl|rl||}
\hline
\hline
\multicolumn{8}{||c||}{ \textbf{Charmonium $0^-$ Spectrum}} \\
\hline
\hline
\multicolumn{4}{||c||}{$\chi_c (1P)$} &
\multicolumn{4}{c||}{$\chi_b (1P)$} \\
\hline
 $1s^2_{1/2}$  &  2721.46 & $\eta_c (1S)$  & $8.7 \%$  &
 $1p^2_{3/2}$  &  3520.13 & $\eta_c (2S)$  & $2.1 \%$  \\
\hline
 $1s_{1/2}; 2s_{1/2}$  &  3415.10 & $\eta_c (2S)$  & $4.98 \%$  &
 $3p^2_{3/2}$  &  3846.38 &  &  \\
\hline
 $1s_{1/2};3s_{1/2}$  &  3583.15 & $\eta_c (2S)$  & $0.30 \%$  &
   &   &   &   \\
\hline
 $2p^2_{1/2}$  &  4108.85 &  &  &
   &   &  &   \\
\hline
\hline
\end{tabular}
\end{center}
\begin{center}
\begin{tabular}[h]{||cl|rl||}
\hline
\hline
\multicolumn{4}{||c||}{ \textbf{Bottomonium $0^-$ Spectrum} ($\chi_b (1P))$} \\
\hline
\hline
 $1s^2_{1/2}$  &  8935.20 & $\eta_b (1S)$  & $3.9 \%$  \\
\hline
 $1s_{1/2};2s_{1/2}$  &  9859.90 &  &  \\
\hline
\hline
\end{tabular}
\end{center}
The tables read as follows (from left to right): configuration, 
theoretical particle mass em MeV, closest observed particle, absolute error in
the theoretical prediction of the particle mass. 
For charmonium the table includes configurations built from
the two independent spectrum. From now on, each particle will appear only
in the line refering to the closest mass prediction in the respective table.
The tables shows that the difference between the theoretical predicitions
and the observed masses is below $10 \%$. The largest error is for 
$1 s^2_{1/2}$ configuration, where one expects to see a significant Lamb shift
correction. Moreover, our description of mesons predicts more states than 
those observed in experiments. As we will see below, this comment applies to 
all particles quantum number.

For $J^P \, = \, 0^+$ particles our predictions are resumed
in the following tables
\begin{center}
\begin{tabular}[h]{||cl|rl||}
\hline
\hline
\multicolumn{4}{||c||}{ \textbf{Charmonium $0^+$ Spectrum} ($\chi_c (1P)$)} \\
\hline
\hline
 $1s_{1/2}; 2p_{1/2}$  &  3415.10 & $\chi_{c0} (1P)$  & $0.0 \%$  \\
\hline
 $1s_{1/2}; 3p_{1/2}$  &  3583.15 &                   &           \\
\hline
\hline
\end{tabular}
\end{center}
\begin{center}
\begin{tabular}[h]{||cr|rl||}
\hline
\hline
\multicolumn{4}{||c||}{ \textbf{Bottomonium $0^+$ Spectrum} ($\chi_b (1P))$} \\
\hline
\hline
 $1s_{1/2}; 2p_{1/2}$  &  9859.90 & $\chi_{b0} (1P)$  & $0.0 \%$  \\
\hline
 $1s_{1/2};3p_{1/2}$  &  10044.12 & $\chi_{b0} (2P)$  & $1.8 \%$  \\
\hline
 $2s_{1/2};2p_{1/2}$  &  10784.63 &   &  \\
\hline
\hline
\end{tabular}
\end{center}
Note that $\chi_0 (1P)$ were used as inputs to compute $E_{glue}$ and
$F$. For $0^+$ states, only the predicition for $\chi_{b0} (2P)$ can be
compared with the observed data. The theoretical value for the mass of
$\chi_{b0} (2P)$ has an error which is below $2 \%$. 
Naivelly, one would expect $s$ states to have larger corrections than $p$ 
states coming from higher order processes. In particular, the Lamb shift
correction is negligable small for a $3p$ state, significant for $1s$ state, 
quite small for $2s$ and smaller for a $3s$ state.
This relative importance of the Lamb shift correction at least points in the
right direction to explain the errors listed in the above tables. 
One should not forget that there are corrections to this
particle-independent description of mesons. The nature of these corrections
is substantially different from the Lamb shift correction and they are
expected to be larger when the overlap between quarks is larger; for example, 
when quark and antiquark are in the same orbital.

For $J^P \, = \, 1^-$ particles our predictions are resumed
in the following tables
\begin{center}
\begin{tabular}[h]{||cr|ll||}
\hline
\hline
\multicolumn{4}{||c||}{ \textbf{Charmonium $1^-$ Spectrum} ($\chi_c (1P)$)} \\
\hline
\hline
 $1s_{1/2}; 2s_{1/2}$  &  3415.10 & $J/ \psi$  & $10.3 \%$  \\
\hline
 $1s_{1/2}; 3s_{1/2}$  &  3583.15 &                   &           \\
\hline
 $1s_{1/2}; 3d_{3/2}$  &  3617.49 &   &   \\
\hline
 $2p_{1/2}; 2p_{3/2}$  &  4227.00 & $\psi (4040)$  & $4.6 \%$  \\
\hline
 $2s_{1/2}; 3s_{1/2}$  &  4276.79 & $\psi (4160)$  & $2.8 \%$  \\
 $2p_{1/2}; 3p_{1/2}$  &  4276.79 &                &           \\
\hline
 $2s_{1/2}; 3d_{3/2}$  &  4276.79 & $\psi (4160)$  & $2.8 \%$  \\
 $2p_{1/2}; 3p_{3/2}$  &  4311.14 &                &           \\
\hline
 $2p_{3/2}; 3p_{1/2}$  &  4395.04 &   &  \\
\hline
 $2p_{3/2}; 3p_{3/2}$  &  4429.39 & $\psi (4415)$  & $0.3 \%$  \\
\hline
 $3p_{1/2}; 3p_{3/2}$  &  4451.06 &  &   \\
\hline
\hline
\end{tabular}
\end{center}
\begin{center}
\begin{tabular}[h]{||cr|ll||}
\hline
\hline
\multicolumn{4}{||c||}{ \textbf{Charmonium $1^-$ Spectrum} ($\chi_b (1P)$)} \\
\hline
\hline
 $2p_{3/2}; 2p_{3/2}$  &  3683.26 & $\psi (2S)$  & $0.07 \%$  \\
\hline
 $3d_{3/2}; 3d_{5/2}$  &  3863.30 & $\psi (3770)$ & $2.5 \%$ \\
\hline
 $n \ge 4$  state &   &   &   \\
\hline
\hline
\end{tabular}
\end{center}
\begin{center}
\begin{tabular}[h]{||cr|rl||}
\hline
\hline
\multicolumn{4}{||c||}{ \textbf{Bottomonium $1^-$ Spectrum} ($\chi_b (1P))$} \\
\hline
\hline
 $1s_{1/2}; 2s_{1/2}$  &  9859.90 & $\Upsilon (1S)$  & $4.2 \%$  \\
\hline
 $1s_{1/2};3s_{1/2}$  &  10044.12 & $\Upsilon (2S)$  & $0.2 \%$  \\
\hline
 $2s_{1/2};3d_{3/2}$  &  10056.87 & $\Upsilon (3S)$  & $2.9 \%$  \\
\hline
 $2p_{1/2};3p_{3/2}$  &  10056.87 & $\Upsilon (4S)$  & $2.3 \%$  \\
\hline
 $2s_{1/2};3s_{1/2}$  &  10968.85 & $\Upsilon (10860)$  & $1.0 \%$  \\
 $2p_{1/2};3p_{1/2}$  &  10968.85 &   &   \\
\hline
 $2s_{1/2};3d_{3/2}$  &  10981.61 &  &  \\
 $2p_{1/2};3p_{3/2}$  &  10981.61 &  &  \\
\hline
 $2p_{3/2};3p_{1/2}$  &  11011.60 &  &   \\
\hline
 $2p_{3/2};3p_{3/2}$  &  11024.36 & $\Upsilon (11020)$  & $0.05 \%$  \\
\hline
 $3p_{1/2};3p_{3/2}$  &  11165.83 &  &  \\
 $3s_{1/2};3d_{3/2}$  &  11165.83 &  &  \\
\hline
\hline
\end{tabular}
\end{center}
Apart $J/ \psi$, the errors in the theoretical predictions are under $5 \%$. 
For $\psi (2S)$, $\psi (4415)$, $\Upsilon (2S)$, $\Upsilon (10860)$ and
$\Upsilon (11020)$ the agreement between theory and experiment is below
$1 \%$. Such an excelent match between theory and experiment seems to suggest
that our picture of hadronic matter can provide a good starting point
for more precise heavy hadron computations. In what concerns $J/ \psi$,
the error in the theoretical mass is the largest in the calculation discussed
in this paper. Again, one should not forget that $J/ \psi$ is described by
a configuration where one expects to see large corrections. Note that
the corresponding state in the bottomonium spectrum $\Upsilon (1S)$ has
the largest error in the bottomonium spectrum. For $J/ \psi$ and
$\Upsilon (1S)$, the error in the theoretical value for their mass 
is roughly a factor of two larger than the next largest error.

For $J^P \, = \, 1^+$ particles our predictions are resumed
in the following tables
\begin{center}
\begin{tabular}[h]{||cr|ll||}
\hline
\hline
\multicolumn{4}{||c||}{ \textbf{Charmonium $1^+$ Spectrum} ($\chi_c (1P)$)} \\
\hline
\hline
 $1s_{1/2}; 2p_{1/2}$  &  3510.51 &  &   \\
\hline
 $1s_{1/2}; 2p_{3/2}$  &  3533.35 &  $\chi_{c1} (1P)$ & $0.7 \%$   \\
\hline
 $1s_{1/2}; 3p_{1/2}$  &  3617.49 &   &   \\
\hline
\hline
\end{tabular}
\end{center}
\begin{center}
\begin{tabular}[h]{||cr|rl||}
\hline
\hline
\multicolumn{4}{||c||}{ \textbf{Bottomonium $1^+$ Spectrum} ($\chi_b (1P))$} \\
\hline
\hline
 $1s_{1/2}; 2p_{1/2}$  &  9859.90 &  &  \\
\hline
 $1s_{1/2}; 2p_{3/2}$  &  9902.65 & $\chi_{b1} (1P)$  & $0.1 \%$  \\
\hline
 $1s_{1/2}; 3p_{1/2}$  &  10044.12 &   &  \\
\hline
 $1s_{1/2}; 3p_{3/2}$  &  10056.87 & $\chi_{b1} (2P)$  & $1.9 \%$  \\
\hline
 $2p_{1/2}; 2p_{3/2}$  &  10827.38 &  &  \\
\hline
\hline
\end{tabular}
\end{center}
Remember that $\chi_1 (1P)$ was used as input to compute $F$. Similarly
as what was observed previously, the only predict mass has an error of 
$\sim 2 \%$.

For $J^P \, = \, 2^+$ particles our predictions are resumed
in the following tables
\begin{center}
\begin{tabular}[h]{||cr|ll||}
\hline
\hline
\multicolumn{4}{||c||}{ \textbf{Charmonium $2^+$ Spectrum} ($\chi_c (1P)$)} \\
\hline
\hline
 $1s_{1/2}; 2p_{3/2}$  &  3533.35 &  $\chi_{c2} (1P)$ & $0.7 \%$   \\
\hline
 $1s_{1/2}; 3p_{1/2}$  &  3617.49 &   &   \\
\hline
\hline
\end{tabular}
\end{center}
\begin{center}
\begin{tabular}[h]{||cr|rl||}
\hline
\hline
\multicolumn{4}{||c||}{ \textbf{Bottomonium $2^+$ Spectrum} ($\chi_b (1P))$} \\
\hline
\hline
 $1s_{1/2}; 2p_{3/2}$  &  9902.65 & $\chi_{b2} (1P)$  & $0.1 \%$  \\
\hline
 $1s_{1/2}; 3p_{3/2}$  &  10056.87 & $\chi_{b2} (2P)$  & $2.1 \%$  \\
\hline
 $2p_{1/2}; 2p_{3/2}$  &  10827.38 &  &  \\
\hline
\hline
\end{tabular}
\end{center}
Remember that $\chi_2 (1P)$ states where used to compute $F$. For the particle
$\chi_{b2} (2P)$, the situation is similar to $\chi_{b1} (2P)$.

For $J^P \, = \, 2^-$ particles our predictions are resumed
in the following tables
\begin{center}
\begin{tabular}[h]{||cr|ll||}
\hline
\hline
\multicolumn{4}{||c||}{ \textbf{Charmonium $2^-$ Spectrum} ($\chi_c (1P)$)} \\
\hline
\hline
 $1s_{1/2}; 3d_{3/2}$  &  3617.49 &   &   \\
\hline
 $1s_{1/2}; 3d_{5/2}$  &  3621.61 &   &   \\
\hline
 $1p_{1/2}; 2p_{3/2}$  &  4227.00 &   &   \\
\hline
\hline
\end{tabular}
\end{center}
\begin{center}
\begin{tabular}[h]{||cr|ll||}
\hline
\hline
\multicolumn{4}{||c||}{ \textbf{Charmonium $2^-$ Spectrum} ($\chi_b (1P)$)} \\
\hline
\hline
 $2p^2_{3/2}$  &  3520.13 &  &    \\
\hline
 $2p_{3/2}; 3p_{3/2}$  &  3683.26 &  &   \\
\hline
 $3p^2_{3/2}$  &  3846.38 & $\psi (3836)$ & $0.3 \%$   \\
 $3d^2_{3/2}$  &  3846.38 &  &    \\
\hline
 $3d_{3/2}; 3d_{5/2}$  &  3863.30 &&   \\
\hline
\hline
\end{tabular}
\end{center}
\begin{center}
\begin{tabular}[h]{||cr|rl||}
\hline
\hline
\multicolumn{4}{||c||}{ \textbf{Bottomonium $2^-$ Spectrum} ($\chi_b (1P))$} \\
\hline
\hline
 $1s_{1/2}; 3d_{3/2}$  &  10056.87 &  &  \\
\hline
 $1s_{1/2}; 3d_{5/2}$  &  10060.14 &  &  \\
\hline
 $2p_{1/2}; 2p_{3/2}$  &  10827.38 &  &   \\
\hline
\hline
\end{tabular}
\end{center}
Unfortunately, only one $2^-$ was observed and we can not say much about
our predictions. For $\psi (3836)$ the theoretical prediction matches the
experimental mass with high accuracy. Note that for the configuration
$3p^2_{3/2}$ the Lamb shift correction is expected to be negligable.

In the charmonium spectrum there is a particle whose quantum numbers are not
known, $h_c (1P)$. Following the same reasoning as before, we can look for the
closest theoretical mass value. For $h_c (1P)$ we find two configurations
which are very good candidates to represent $h_c (1P)$, namely
\begin{eqnarray}
  \left[ \chi_c (1P) \right] ~~1s_\frac{1}{2}; 2p_\frac{3}{2} (1^+, 2^+) & &
   M \, = \,  3533.35 \mbox{ MeV  -    Error = }0.2 \% \, ,\\
  \left[ \chi_b (1P) \right] ~~ 2p^2_\frac{3}{2} (0^-, 2^-) & &
   M \, = \,  3520.13 \mbox{ MeV  -    Error = }0.2 \% \, .
\end{eqnarray}

In what concerns the charmonium and bottomonium spectrum, one can conclude
that the particle-independent description is able to reproduce the particle 
masses with a $10 \%$ precision. Indeed, for almost all
particles, the level of precision is below $5 \%$ error and for same 
particles, the level of precision is under $1 \%$. This result is encouraging,
specially if one notes that for these last states no large corrections, from
higher order processes, are expected to take place.
Moreover, the particle-independent picture relates the two spectrum and 
predicts a number of new particles that can be tested experimentally.
Another interesting observation being that what we call $E_{glue}$, see
(\ref{eglue}) for values, is compatible with the quenched lattice QCD
estimates for the ground state and first excited state $0^+$ glueballs 
\cite{Mi03}.

The particle data book \cite{pdg02} reports a bottom-charm meson with a mass
of $6.4 \, \pm 0.39 \, \pm 0.13$ GeV. There is not to much  experimental
information on the properties of this meson. For example, the quantum numbers
of this particle were never measured. Assuming, that the bottom-charm meson is
the lowest mass state, our picture of a meson gives the following prediction
\begin{eqnarray}
  &  &  6.71 \mbox{ GeV}, ~ J^P \, = \,~ 0^-, \, 1^-
                        \mbox{ from } \chi_c (1P) \mbox{ spectrum } \\
  &  &  6.23 \mbox{ GeV}, ~ J^P \, = \,  1^+, \, 2^+  ~
                        \mbox{ from } \chi_b (1P) \mbox{ spectrum} .
   \label{bcmeson}
\end{eqnarray}
The best agreement between theory and experiment happens for (\ref{bcmeson})
state (error  is $2.7 \%$). 

For baryons the particle-independent model also makes predictions. 
Unfortunately, there is no experimental information on baryons made up
of only $c$ or $b$ quarks. Instead of giving a detailed spectrum, we
simply quote the prediction for the lowest $bbb$ and $ccc$ states
\begin{eqnarray}
 M \left( bbb \right) ~ = ~ 12.54 ~ \mbox{GeV} &  & J^P ~ = ~ \frac{3}{2}^+
        \, ,
\\
 M \left( ccc \right) ~ = ~ 2.96 ~ \mbox{GeV} &  & J^P ~ = ~ \frac{3}{2}^+
        \, .
\end{eqnarray} 
Surprisingly, the charm predicition is a rather light state.

\section{Results and Conclusions}

In this paper we report on Minkowsky space-time solutions of classical pure 
SU(2) and SU(3) gauge theories in Landau gauge. The solutions were obtained 
after writing the gluon field in a particular way (\ref{A}) and choosing a 
spherical like basis in color space. The two steps seem to be crucial to map 
SU(2) and SU(3) to abelian like theories, replacing the nonlinear classical
field equations by linear equations. If the construction of the ansatz
remembers the ``abelian projection'' technique, in our case, it is only after
the choice of the spherical like basis that the ``abelian nature'' of the 
theory shows up. On the other hand, the ansatz transforms the Landau gauge 
condition, a linear gauge fixing condition, in a set of coupled equations.

Despite same appealing properties of the ansatz (\ref{A}), we do not know if
in general our gluon field provides a good dynamical description of the
gluonic degrees of freedom. If so, does it means that the dynamics of 
classical gauge theories is, essentially, the dynamics of various uncoupled 
photonic fields? The idea is interesting but, at present, we cannot answer 
this question.

The classical configurations discussed are regular everywhere except at the
origin and/or infinity. Moreover, the SU(2) and SU(3) solutions of the
Euler-Lagrange equations are similar to classical electrodynamic fields.
The difference is a longitudinal component in the gauge fields, which
does not contributes neither to the energy nor the spin of the theory but 
couples to the fermionic fields. Like in classical electrodynamics, the 
solutions are exponential functions and are identified by a light-type 
four-vector $k_\mu$, $k^2 \, = \, 0$. The longitudinal component of the gauge 
field is given by a combination of pure complex exponential functions and real
exponential functions. These last components make the gluonic field to diverge
for large space and time values.

In what concerns the coupling of the classical configurations to fermionic
fields, the quark-gluon interaction is equivalent to three independent 
electro\-ma\-gne\-tic\--\-type couplings. 
The solution of the Dirac equation for quarks in the background of the gluon
fields computed in section 4, suggests that physical particles are the
usual quark model mesons and baryons. The hadronic wave function is given by
the slater determinant of single-quark wave functions for electromagnetic 
problems. In this picture, an hadron is described by a particle-independent
model like in Atomic and Nuclear Physics.

The classical gluonic configurations of section 4 include, as particular 
solution, the Coulomb potential. The spectrum of the Coulomb problem is well 
known. If one assumes a particle-independent model, for this particular 
solution it is possible to compute the associated particle spectrum. For 
heavy quarkonium our calculation shows that such a picture reproduces the
full quarkonium spectrum with an error of $10 \%$ or bellow. Indeed, typicall 
errors in mass prediction are between $2 \%$ and $5 \%$ and for same states 
the error in the theoretical mass is bellow $1 \%$. Note that, if the
Coulomb gluonic configurations are the relevant configurations for heavy 
quarkonium, such an approach is similar to the quenched approximation used
in lattice QCD. One should not forget that such a picture of what is an hadron
is not the full story but, maybe, a good starting point for a more precise
and detailed calculation.

The result for the charmonium and bottomonium spectrum seems to suggest that
the classical gluonic configurations computed here can tell us something about
hadronic properties. However, before giving a clear answer to such a question
a number of other 
questions have to be answered: can the coulomb picture described
in this paper reproduce the meson properties (masses, decay widths, etc.) to
a high precision? What is the meaning of $E_{glue}$? Should one think on
mesons as a $q {\overline q} \otimes \left( 0^+ ~ \mbox{glueball} \right)$?
What about light mesons and baryons?

\section*{Acknowledge}

R. A. Coimbra acknowledges the financial support from FCT, grant
BD/8736/2002.

\section*{Appendix}

The color space has dimension eight. For an eight dimension space, the
spherical basis requires seven angles. Considering the following seven
functions $\theta_i (x)$, $i = 1 ... 7$ we define the basis as follows
\begin{eqnarray}
 \vec{e}_1 \, = \,  & & \left(\begin{array}{l}
     \sin \theta_1 \, \cos \theta_2 \, \sin \theta_3 \, \sin \theta_4 \,
                       \sin \theta_5 \, \sin \theta_6 \, \sin \theta_7  \\
     \sin \theta_1 \, \sin \theta_2 \, \sin \theta_3 \, \sin \theta_4  \,
                       \sin \theta_5 \, \sin \theta_6 \, \sin \theta_7  \\
     \cos \theta_1 \, \sin \theta_3 \, \sin \theta_4 \,
                       \sin \theta_5 \, \sin \theta_6 \, \sin \theta_7  \\
     \cos \theta_3 \, \sin \theta_4 \,
                       \sin \theta_5 \, \sin \theta_6 \, \sin \theta_7  \\
     \cos \theta_4 \,  \sin \theta_5 \, \sin \theta_6 \, \sin \theta_7  \\
     \cos \theta_5 \, \sin \theta_6 \, \sin \theta_7   \\
     \cos \theta_6 \, \sin \theta_7  \\
     \cos \theta_7
                            \end{array}  \right) \\
 \vec{e}_2 \, = \,  & & \left(\begin{array}{l}
     \cos \theta_1 \, \cos \theta_2  \\
     \cos \theta_1 \, \sin \theta_2  \\
     - \sin \theta_1                 \\
      0                              \\
      0                              \\
      0                              \\
      0                              \\
      0         
                            \end{array}  \right) \\
 \vec{e}_3 \, = \,  & & \left(\begin{array}{l}
     - \sin \theta_2                 \\
     \cos \theta_2                   \\
     0                               \\
      0                              \\
      0                              \\
      0                              \\
      0                              \\
      0         
                            \end{array}  \right) \\
 \vec{e}_4 \, = \,  & & \left(\begin{array}{l}
     \sin \theta_1 \, \cos \theta_2 \, \cos \theta_3  \\
     \sin \theta_1 \, \sin \theta_2 \, \cos \theta_3  \\
     \cos \theta_1 \, \cos \theta_3                   \\
      - \sin \theta_3                              \\
      0                              \\
      0                              \\
      0                              \\
      0         
                            \end{array}  \right) \\
 \vec{e}_5 \, = \,  & & \left(\begin{array}{l}
     \sin \theta_1 \, \cos \theta_2 \, \sin \theta_3 \, \cos \theta_4 \\
     \sin \theta_1 \, \sin \theta_2 \, \sin \theta_3 \, \cos \theta_4 \\
     \cos \theta_1 \, \sin \theta_3 \, \cos \theta_4                   \\
     \cos \theta_3 \, \cos \theta_4                              \\
      - \sin \theta_4                              \\
      0                              \\
      0                              \\
      0                             \end{array}  \right) \\
 \vec{e}_6 \, = \,  & & \left(\begin{array}{l}
     \sin \theta_1 \, \cos \theta_2 \, \sin \theta_3 \, \sin \theta_4
                   \, \cos \theta_5 \\
     \sin \theta_1 \, \sin \theta_2 \, \sin \theta_3 \, \sin \theta_4 
                   \, \cos \theta_5 \\
     \cos \theta_1 \, \sin \theta_3 \, \sin \theta_4 \, \cos \theta_5   \\
     \cos \theta_3 \, \sin \theta_4 \, \cos \theta_5                \\
     \cos \theta_4 \, \cos \theta_5                              \\
      - \sin \theta_5                              \\
      0                              \\
      0                             \end{array}  \right) \\
 \vec{e}_7 \, = \,  & & \left(\begin{array}{l}
     \sin \theta_1 \, \cos \theta_2 \, \sin \theta_3 \, \sin \theta_4
                   \, \sin \theta_5 \, \cos \theta_6 \\
     \sin \theta_1 \, \sin \theta_2 \, \sin \theta_3 \, \sin \theta_4 
                   \, \sin \theta_5 \, \cos \theta_6 \\
     \cos \theta_1 \, \sin \theta_3 \, \sin \theta_4 \, \sin \theta_5  
                   \, \cos \theta_6 \\
     \cos \theta_3 \, \sin \theta_4 \, \sin \theta_5 \, \cos \theta_6     \\
     \cos \theta_4 \, \sin \theta_5 \, \cos \theta_6            \\
     \cos \theta_5 \, \cos \theta_6                              \\
      - \sin \theta_6                              \\
      0                             \end{array}  \right) \\
 \vec{e}_8 \, = \,  & & \left(\begin{array}{l}
     \sin \theta_1 \, \cos \theta_2 \, \sin \theta_3 \, \sin \theta_4
                   \, \sin \theta_5 \, \sin \theta_6 \, \cos \theta_7 \\
     \sin \theta_1 \, \sin \theta_2 \, \sin \theta_3 \, \sin \theta_4 
                   \, \sin \theta_5 \, \sin \theta_6 \, \cos \theta_7 \\
     \cos \theta_1 \, \sin \theta_3 \, \sin \theta_4 \, \sin \theta_5  
                   \, \sin \theta_6 \, \cos \theta_7 \\
     \cos \theta_3 \, \sin \theta_4 \, \sin \theta_5 \, \sin \theta_6 
                   \, \cos \theta_7    \\
     \cos \theta_4 \, \sin \theta_5 \, \sin \theta_6 \, \cos \theta_7   \\
     \cos \theta_5 \, \sin \theta_6 \, \cos \theta_7                \\
     \cos \theta_6 \, \cos \theta_7                             \\
      - \sin \theta_7                           \end{array}  \right)
\end{eqnarray}



\begin{thebibliography}{99}

\bibitem{Ac79} A. Actor, \rmp{51}{1979}{461}

\bibitem{Ho79} G. 't Hooft, \nucl{B153}{1979}{141}

\bibitem{Ho81} G. 't Hooft, \nucl{B190[FS]}{1981}{455}

\bibitem{Po77} A. Polyakov, \nucl{B120}{1977}{429}

\bibitem{ScSh98} T. Schafer, E. V. Shuryak, \rmp{70}{1998}{323}

\bibitem{Di02} D. Diakonov, hep-th/0212026.

\bibitem{Ch991} Y. M. Cho, \prd{62}{2000}{074009}

\bibitem{Ch992} Y. M. Cho, J. Korean Phys. Soc. \textbf{38} (2001) 15 and
hep-th/9906198.

\bibitem{ChLe99} Y. M. Cho, H. Lee, D. G. Pak, hep-th/9905215.

\bibitem{FaNi99} L. Faddeev, A. J. Niemi, \plb{464}{1999}{90}

\bibitem{eu} O. Oliveira, hep-th/0105222. See also \cite{Li00}.

\bibitem{Li00} S. Li, Y. Zhang, Z. Zhu, \plb{487}{2000}{201}

\bibitem{Neu94} M. Neubert, \prep{245}{1994}{259}

\bibitem{KP94} J. G. K\"orner, D. Pirjol, M. Kr\"ammer, \ppnp{33}{1994}{787}

\bibitem{LP96} See, for example, E. Leader, E. Predazzi,
\textit{An introduction to gauge theories and modern particle physics},
Cambridge University Press, 1996.

\bibitem{MM94} See, for example, I. Montvay, G. M\"unster,
 \textit{Quantum Fields on a Lattice}, Cambridge University Press, 1994.

\bibitem{IZ80} See, for example, C. Itzykson, J. B. Zuber,
\textit{Quantum Field Theory}, McGraw-Hill Book Co, 1985.

\bibitem{pdg02} K. Hagiwara \textit{et al.}, \prd{66}{2002}{010001-1}

\bibitem{alphaS} http://www-theory.lbl.gov/~ianh/alpha/alpha.html

\bibitem{Mi03} C. Michael, in 
\textit{International Review of Nuclear Physics}, Vol. 9, 
\textit{Hadronic Physics from Lattice QCD}, edited by A. M. Green, World
Scientific and hep-lat/0302001


\end{thebibliography}
\end{document}